\documentclass[aps,prd,twocolumn,showkeys,amsmath,amssymb]{revtex4}
\usepackage{amsfonts}
\usepackage{dcolumn}
\usepackage{bm}
\usepackage{txfonts}
\usepackage{feynmf}   %{feynmp}
\usepackage{slashed}  %for Feynman symbols

\usepackage{graphicx}
\usepackage{graphics}
\usepackage{subfigure}
\usepackage{epstopdf}
\usepackage{epsfig}
\usepackage{overpic}

\usepackage{booktabs}
\usepackage{float}
\usepackage{longtable,lscape}
\usepackage{indentfirst}
\usepackage{dashrule}
\usepackage{multirow}
\pdfoutput=1

\usepackage[colorlinks,citecolor=blue,anchorcolor=red,menucolor=red,linkcolor=red,filecolor=red,runcolor=red,urlcolor=blue,frenchlinks=red]{hyperref}

\allowdisplaybreaks[3]

\newcounter{cnt}
\makeatletter

\let\oldhypertarget\hypertarget
\renewcommand{\hypertarget}[2]{%
  \oldhypertarget{#1}{#2}%
    \protected@write\@mainaux{}{%
        \string\expandafter\string\gdef
          \string\csname\string\detokenize{#1}\string\endcsname{#2}%
    }%
  }
\newcommand{\myhyperlink}[1]{%
  \hyperlink{#1}{\csname #1\endcsname}%
  }
\makeatother

\newcommand{\cref}[1]{\myhyperlink{#1}}

\renewcommand{\arraystretch}{1.8}
\begin{document}

\title{Investigating charmed hybrid baryons via QCD sum rules}

\author{Hui-Min Yang$^1$}
\author{Xuan Luo$^2$}
\author{Hua-Xing Chen$^2$}
\email{hxchen@seu.edu.cn}
\author{Wei Chen$^{3,4}$}
\email{chenwei29@mail.sysu.edu.cn}

\affiliation{
$^1$School of Physics and Center of High Energy Physics, Peking University, Beijing 100871, China \\
$^2$School of Physics, Southeast University, Nanjing 210094, China\\
$^3$School of Physics, Sun Yat-Sen University, Guangzhou 510275, China\\
$^4$Southern Center for Nuclear-Science Theory (SCNT), Institute of Modern Physics, Chinese Academy of Sciences, Huizhou 516000, Guangdong province, China
}

\begin{abstract}
We investigate charmed hybrid baryons using the QCD sum rule method within the framework of heavy quark effective theory. We construct twenty-eight interpolating currents for charmed hybrid baryons, seven of which are employed in QCD sum rule analyses of nineteen states with quark-gluon configurations $qqcg$, $qscg$, and $sscg$ ($q = u/d$). The masses of the lowest-lying charmed hybrid baryons in the $SU(3)$ flavor $\mathbf{6}_F$ representation are calculated to be $M_{\Sigma_{cg}(1/2^+)} =  3.36^{+0.27}_{-0.26}~\rm{GeV}$, $M_{\Xi^\prime_{cg}(1/2^+)} = 3.59\pm 0.20~\rm{GeV}$, and
$M_{\Omega_{cg}(1/2^+)} = 3.82\pm 0.21~\rm{GeV}$. We propose that future experiments search for these states via their $P$-wave decay channels $ND^{(*)}$, $\Lambda D^{(*)}$, and $\Xi D^{(*)}$, respectively. Such investigations would provide valuable insight into the role of gluonic excitations in hadron structure.
\end{abstract}


\maketitle

\section{Introduction}

Quantum Chromodynamics (QCD) allows for the existence of color singlets, including conventional hadrons, multi-quark states, glueballs, and hybrids. Exploring exotic states beyond conventional hadrons is one of the most intriguing areas of research in hadronic physics, as such studies enhance our understanding of non-perturbative QCD. Gluonic degrees of freedom play a crucial role in these exotic states. The excitation of the gluonic field, analogous to the role of valence quarks, manifests in glueballs, hybrid mesons, and hybrid baryons. While substantial research has been conducted on hybrid mesons, relatively few studies have focused on hybrid baryons. Consequently, investigating hybrid baryons beyond conventional hadron configurations remains a significant and challenging topic in hadronic physics, requiring the identification of distinctive characteristics to confirm their existence. A common approach involves predicting the mass spectra and decay widths of hybrid baryons and offering specific recommendations for experimental searches.

In 2022 the BESIII collaboration observed the isoscalar state $\eta_1(1855)$ with the exotic quantum numbers $I^G J^{PC} = 0^+ 1^{-+}$~\cite{BESIII:2022riz,BESIII:2022iwi}. Additionally, several collaborations~\cite{IHEP-Brussels-LosAlamos-AnnecyLAPP:1988iqi,E852:1997gvf,E852:1998mbq,Aoyagi:1993kn,CrystalBarrel:1998cfz} have reported three isovector states---$\pi_1(1400)$, $\pi_1(1600)$, and $\pi_1(2015)$---all possessing the exotic quantum numbers $I^G J^{PC} = 1^- 1^{-+}$. These states are regarded as strong candidates for hybrid mesons. In 2024, the BESIII collaboration also reported the $X(2370)$ as a possible lightest pseudoscalar glueball~\cite{BESIII:2023wfi,Huang:2025pyv}. These observations collectively support the potential existence of hybrid baryons, although no hybrid hadron has yet been conclusively identified in experiments. Ongoing efforts to search for hybrid baryons are underway at CLAS in Hall B of Jefferson Lab~\cite{Lanza:2017qel,Burkert:2018nvj}, with the first results expected in the coming years. The hybrid hadron system has also attracted considerable interest from theorists. These exotic hadrons have been studied using various theoretical frameworks, including the bag model~\cite{Kusaka:1990ha,Barnes:1977hg,Barnes:1982fj,Barnes:1983xx,Duck:1983ju,Carlson:1983xr}, lattice QCD~\cite{Dudek:2012ag,Bali:1993fb,Chen:2005mg,Morningstar:1999rf,Gui:2019dtm,Michael:1985ne,Juge:2002br,Lacock:1996ny,Hedditch:2005zf,Bernard:2003jd}, the large-$N_c$ approach~\cite{Chow:1998tq}, the flux-tube model~\cite{Page:2004gv,Capstick:2002wm,Capstick:1999qq,Page:1998qt,Isgur:1983wj,Qiu:2022ktc}, the hybrid quark model~\cite{Khosonthongkee:2014gla,Galata:2014typ,Li:1991sh}, the constituent gluon model~\cite{Horn:1977rq,Szczepaniak:2001rg,Guo:2007sm}, and QCD sum rules~\cite{Zhao:2023imq,Kisslinger:1995yw,Kisslinger:2003hk,Azizi:2017xyx,Tan:2024grd,Lian:2024fsb}, among others. Additional discussions on related developments can be found in Refs.~\cite{Cimino:2024bol,Khan:2020ahz,Kittel:2005jm,Page:2002mt,Barnes:2000vn,Chen:2022asf}, which provide insights into recent progress. Notably, most studies of hybrid baryons have concentrated on systems composed entirely of either light or heavy quarks. In contrast, our focus is on singly heavy hybrid baryons containing only one heavy quark. In such systems, the heavy quark effective theory (HQET) can be effectively applied, providing a promising framework for advancing our understanding of non-perturbative QCD.

In this paper we systematically investigate charmed hybrid baryons using the QCD sum rule method within the framework of heavy quark effective theory. We construct twenty-eight interpolating currents for charmed hybrid baryons, seven of which are employed to perform sum rule analyses for nineteen states with quark-gluon configurations $qqcg$, $qscg$, and $sscg$ ($q = u/d$). We carry out comprehensive analyses of their mass spectra for quantum numbers $J^P = 1/2^+$, $3/2^+$, and $5/2^+$, and determine their decay constants, which are essential for understanding their decay behaviors in future studies. In particular, we calculate the masses of the lowest-lying charmed hybrid baryons within the $SU(3)$ flavor $\mathbf{6}_F$ representation to be:
\begin{eqnarray}
\nonumber M_{\Sigma_{cg}(1/2^+)}&=& 3.36^{+0.27}_{-0.26}~\rm{GeV}\, ,
\\ \nonumber M_{\Xi^\prime_{cg}(1/2^+)}&=& 3.59\pm 0.20~\rm{GeV}\, ,
\\ \nonumber M_{\Omega_{cg}(1/2^+)}&=& 3.82\pm 0.21~\rm{GeV}\, .
\end{eqnarray}
All these states carry the quantum numbers $J^P = 1/2^+$. We propose that future experimental studies search for them through their $P$-wave decay channels $ND^{(*)}$, $\Lambda D^{(*)}$, and $\Xi D^{(*)}$, respectively. Such investigations would be instrumental in deepening our understanding of gluonic field excitations within hadronic systems.

This paper is organized as follows. In Sec.~\ref{sec:current} we systematically construct the interpolating currents for charmed hybrid baryons. These currents are employed in Sec.~\ref{sec:leading} to perform QCD sum rule analyses at leading order, and in Sec.~\ref{sec:nexttoleading} to incorporate ${\mathcal O}(1/m_Q)$ corrections. A summary and discussion of our findings are presented in Sec.~\ref{sec:summary}.

\section{Charmed hybrid baryon currents}
\label{sec:current}

In this section we systematically construct the interpolating currents for charmed hybrid baryons, where the heavy quark symmetry plays a crucial role. In the charmed hybrid baryon system, the heavy quark serves as a static color source, and the system's properties are primarily determined by the light degrees of freedom. Let us focus on the ground-state $Qq_1q_2g$ system, where $Q$ represents the heavy $charm$ quark, $q_1$ and $q_2$ represent the two valence light quarks ($up$, $down$, or $strange$), and $g$ represents the valence gluon. In the heavy quark limit, the total angular momentum of the system reduces to
\begin{equation}
J=s_Q\otimes s_{q_1}\otimes s_{q_2}\otimes s_g= s_Q\otimes j_l \, ,
\end{equation}
where $s_Q$, $s_{q_1}$, $s_{q_2}$, and $s_g$ denote the spins of the heavy quark $Q$, the light quarks $q_1$ and $q_2$, and the gluon $g$, respectively. Additionally, $j_l$ denotes the total angular momentum of the light degrees of freedom, expressed as
\begin{equation}
j_l=s_{q_1}\otimes s_{q_2}\otimes s_g= j_{qq}\otimes s_g \, .
\end{equation}
Here, $j_{qq}=s_{q_1}\otimes s_{q_2}$ denotes the combined spin of the two light quarks. These two light quarks satisfy the exchange antisymmetry requirement for identical particles:
\begin{itemize}

\item The flavor structure of the two light quarks can be either antisymmetric or symmetric. Accordingly, we use $\Lambda_{cg}$ and $\Xi_{cg}$ to denote the charmed hybrid baryons in the $SU(3)$ flavor $\bar{\textbf{3}}_F$ representation, while $\Sigma_{cg}$, $\Xi^\prime_{cg}$, and $\Omega_{cg}$ denote those in the $SU(3)$ flavor ${\textbf{6}}_F$ representation.

\item The color structure of the two light quarks can be either antisymmetric $(\bar{\textbf{3}}_C)$ or symmetric $(\textbf{6}_C)$, while the overall color structure of the three quarks is always $\textbf{8}_C$.

\item The spin structure of the two light quarks can be either antisymmetric ($j_{qq}=0$) or symmetric ($j_{qq}=1$).

\end{itemize}

\begin{figure*}[hbtp]
\begin{center}
\scalebox{0.38}{\includegraphics{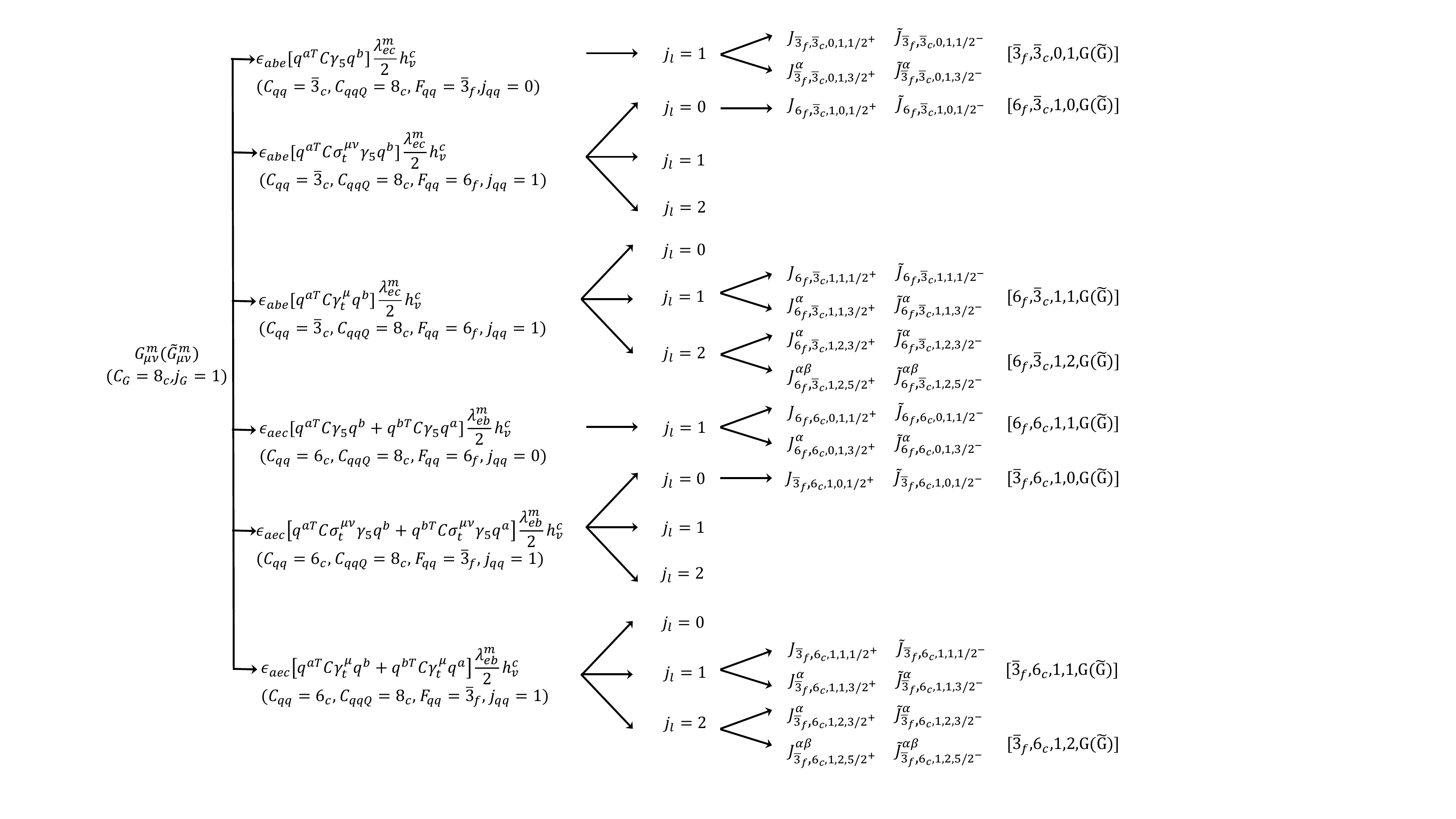}}
\end{center}
\caption{Categorization of charmed hybrid baryon currents.}
\label{fig:hybrid}
\end{figure*}

The charmed hybrid baryon currents can generally be constructed using one charm quark field $h_v^a(x)$, two light quark fields $q_a(x)$ and $q_b(x)$, and one gluon field strength tensor $G_{\mu\nu}^n(x)$ (or one dual gluon field strength tensor $\widetilde{G}_{\mu\nu}^n(x) =G^{n,\rho\sigma}(x) \times \epsilon_{\mu\nu\rho\sigma}/2$), with the color indices $a/b=1\cdots 3$ and $n=1\cdots 8$ as well as the Lorentz indices $\mu/\nu=0\cdots 3$. The (dual) gluon field is a color octet, so the combination of the three quark fields should also form a color octet, with two possible combinations:
\begin{gather}
\epsilon_{abe}[q^{a}q^b]{\lambda_{ec}^n\over 2}h_v^c ~\times~ G^n/\widetilde{G}^n \, ,
\\  
\epsilon_{aec}[q^{a}q^b+q^{b}q^a]{\lambda_{eb}^n\over 2}h_v^c ~\times~ G^n/\widetilde{G}^n \, .
\end{gather}
The light diquark in the former combination is a color anti-triplet, while the one in the latter is a color sextet.

As summarized in Fig.~\ref{fig:hybrid}, we identify a total of twenty-eight charmed hybrid baryon currents, denoted as $J^{\alpha_1\cdots \alpha_{J-1/2}}_{F,C,j_{qq},j_l,J^P}/\widetilde{J}^{\alpha_1\cdots \alpha_{J-1/2}}_{F,C,j_{qq},j_l,J^P}$, which can be categorized into eight multiplets, denoted as $[F,C, j_{qq}, j_l, G/\widetilde{G}]$. Here, $F$ and $C$ represent the flavor and color representations of the two light quarks, while $J(\widetilde{J})$ and $G(\widetilde{G})$ indicate the presence of a gluon (or dual gluon) in the current. The explicit expressions of these currents are given as follows:
\begin{itemize}

\item The doublet $[\bar{3}_f,\bar{3}_c,0,1, G]$ contains two states with $J^P=1/2^+$ and $3/2^+$. Its dual doublet $[\bar{3}_f,\bar{3}_c,0,1, \widetilde{G}]$ contains two other states with $J^P=1/2^-$ and $3/2^-$. Their corresponding currents are
\begin{eqnarray}
J_{\bar{3}_f,\bar{3}_c,0,1,1/2^+} &=& \epsilon_{abe}[q^{aT}\mathcal{C}\gamma_5 q^{b}] \sigma_t^{\mu\nu} g_s {\lambda_{ec}^n\over2}G_{\mu\nu}^n h_v^c\, , 
\label{eq:start}
\\  
\widetilde{J}_{\bar{3}_f,\bar{3}_c,0,1,1/2^-} &=& \epsilon_{abe}[q^{aT}\mathcal{C}\gamma_5 q^{b}] \sigma_t^{\mu\nu} g_s {\lambda_{ec}^n\over2}\widetilde{G}_{\mu\nu}^n h_v^c\, ,
\\ \nonumber
J_{\bar{3}_f,\bar{3}_c,0,1,3/2^+}^\alpha &=& \epsilon_{abe}[q^{aT}\mathcal{C}\gamma_5 q^{b}]
\\ &\times& (g_t^{\alpha\mu}-{1\over3}\gamma_t^\alpha\gamma_t^\mu) \gamma_t^\nu \gamma_5 g_s  {\lambda_{ec}^n\over2}G_{\mu\nu}^n h_v^c\, ,
\\ \nonumber
\widetilde{J}_{\bar{3}_f,\bar{3}_c,0,1,3/2^-}^\alpha &=& \epsilon_{abe}[q^{aT}\mathcal{C}\gamma_5 q^{b}]
\\ &\times& (g_t^{\alpha\mu}-{1\over3}\gamma_t^\alpha\gamma_t^\mu) \gamma_t^\nu \gamma_5 g_s  {\lambda_{ec}^n\over2}\widetilde{G}_{\mu\nu}^n h_v^c\, .
\end{eqnarray}

\item The singlet $[6_f,\bar{3}_c,1,0,G]$ contains one state with $J^P=1/2^+$, and its dual singlet $[6_f,\bar{3}_c,1,0,\widetilde{G}]$ contains another state with $J^P=1/2^-$. Their corresponding currents are
\begin{eqnarray}
J_{6_f,\bar{3}_c,1,0,1/2^+} &=& \epsilon_{abe}[q^{aT}\mathcal{C}\sigma_t^{\mu\nu} \gamma_5 q^{b}] g_s {\lambda_{ec}^n\over2 }G_{\mu\nu}^n h_v^c\, ,
\\ 
\widetilde{J}_{6_f,\bar{3}_c,1,0,1/2^-} &=& \epsilon_{abe}[q^{aT}\mathcal{C}\sigma_t^{\mu\nu} \gamma_5 q^{b}] g_s {\lambda_{ec}^n\over2}\widetilde{G}_{\mu\nu}^n h_v^c\, .
\end{eqnarray}

\item The doublet $[6_f,\bar{3}_c,1,1, G]$ contains two states with $J^P=1/2^+$ and $3/2^+$. Its dual doublet $[6_f,\bar{3}_c,1,1, \widetilde{G}]$ contains two other states with $J^P=1/2^-$ and $3/2^-$. Their corresponding currents are
\begin{eqnarray}
J_{6_f,\bar{3}_c,1,1,1/2^+} &=& \epsilon_{abe}[q^{aT}\mathcal{C}\gamma_t^\mu q^{b}]\gamma_t^{\nu} \gamma_5 g_s {\lambda_{ec}^n\over2}G_{\mu\nu}^n h_v^c\, ,
\\  
\widetilde{J}_{6_f,\bar{3}_c,1,1,1/2^-} &=& \epsilon_{abe}[q^{aT}\mathcal{C}\gamma_t^\mu q^{b}]\gamma_t^{\nu} \gamma_5 g_s {\lambda_{ec}^n\over2}\widetilde{G}_{\mu\nu}^n h_v^c\, ,
\\ \nonumber 
J_{6_f,\bar{3}_c,1,1,3/2^+}^{\alpha} &=& \epsilon_{abe}[q^{aT}\mathcal{C}\gamma_t^\mu q^{b}]
\\ && ~~~~ \times~~ (g^{\alpha\nu}_t-{1\over3}\gamma_t^{\alpha}\gamma_t^\nu) g_s {\lambda_{ec}^n\over2}G_{\mu\nu}^n h_v^c\, ,
\\ \nonumber 
\widetilde{J}_{6_f,\bar{3}_c,1,1,3/2^-}^{\alpha} &=& \epsilon_{abe}[q^{aT}\mathcal{C}\gamma_t^\mu q^{b}]
\\ && ~~~~ \times~~ (g^{\alpha\nu}_t-{1\over3}\gamma_t^{\alpha}\gamma_t^\nu) g_s {\lambda_{ec}^n\over2}\widetilde{G}_{\mu\nu}^n h_v^c\, .
\end{eqnarray}

\item The doublet $[6_f,\bar{3}_c,1,2,G]$ contains two states with $J^P=3/2^+$ and $5/2^+$. Its dual doublet $[6_f,\bar{3}_c,1,2, \widetilde{G}]$ contains two other states with $J^P=3/2^-$ and $5/2^-$. Their corresponding currents are
\begin{eqnarray}
\nonumber J_{6_f,\bar{3}_c,1,2,3/2^+}^{\alpha} &=& \epsilon_{abe}[q^{aT}\mathcal{C}\gamma_{t,\mu_1} q^{b}]
\\ &\times& \Gamma_{J=2}^{\alpha\mu_4,\mu_1\nu_2} \gamma_t^{\mu_4}\gamma_t^{\nu_1} g_s {\lambda_{ec}^n\over2}G_{\nu_1\nu_2}^n h_v^c\, ,
\\ \nonumber
\widetilde{J}_{6_f,\bar{3}_c,1,2,3/2^-}^{\alpha} &=& \epsilon_{abe}[q^{aT}\mathcal{C}\gamma_{t,\mu_1} q^{b}]
\\ &\times& \Gamma_{J=2}^{\alpha\mu_4,\mu_1\nu_2} \gamma_t^{\mu_4}\gamma_t^{\nu_1} g_s {\lambda_{ec}^n\over2}\widetilde{G}_{\nu_1\nu_2}^n h_v^c\, ,
\\ \nonumber
J_{6_f,\bar{3}_c,1,2,5/2^+}^{\alpha\beta} &=& \epsilon_{abe}[q^{aT}\mathcal{C}\gamma_{t,\mu_1} q^{b}]
\\ &\times& \Gamma_{J={5\over 2}}^{\alpha\beta,\mu_1\nu_2} \gamma_t^{\nu_1}\gamma_5 g_s {\lambda_{ec}^n\over2}G_{\nu_1\nu_2}^n h_v^c\, ,
\\ \nonumber
\widetilde{J}_{6_f,\bar{3}_c,1,2,5/2^-}^{\alpha\beta} &=& \epsilon_{abe}[q^{aT}\mathcal{C}\gamma_{t,\mu_1} q^{b}]
\\ &\times& \Gamma_{J={5\over 2}}^{\alpha\beta,\mu_1\nu_2} \gamma_t^{\nu_1}\gamma_5 g_s {\lambda_{ec}^n\over2}\widetilde{G}_{\nu_1\nu_2}^n h_v^c\, .
\label{eq:end}
\end{eqnarray}

\item The doublet $[6_f,6_c,0,1,G]$ contains two states with $J^P=1/2^+$ and $3/2^+$. Its dual doublet $[6_f,6_c,0,1,\widetilde{G}]$ contains two other states with $J^P=1/2^-$ and $3/2^-$. Their corresponding currents are
\begin{eqnarray}
\nonumber J_{6_f,6_c,0,1,1/2^+} &=& \epsilon_{aec}[q^{aT}\mathcal{C}\gamma_5 q^{b}+q^{bT}\mathcal{C}\gamma_5 q^{a}] 
\\ && ~~~~~~~~~~~~~~~~~~~~ \times~~ \sigma_t^{\mu\nu} g_s {\lambda_{eb}^m\over2}G_{\mu\nu}^m h_v^c\, ,
\label{eq:start1}
\\ \nonumber
\widetilde{J}_{6_f,6_c,0,1,1/2^-} &=& \epsilon_{aec}[q^{aT}\mathcal{C}\gamma_5 q^{b}+q^{bT}\mathcal{C}\gamma_5 q^{a}]
\\ && ~~~~~~~~~~~~~~~~~~~~ \times~~ \sigma_t^{\mu\nu} g_s {\lambda_{eb}^m\over2}\widetilde{G}_{\mu\nu}^m h_v^c\, ,
\\ \nonumber
J_{6_f,6_c,0,1,3/2^+}^\alpha &=& \epsilon_{aec}[q^{aT}\mathcal{C}\gamma_5 q^{b}+q^{bT}\mathcal{C}\gamma_5 q^{a}]
\\  &\times& (g_t^{\alpha\mu}-{1\over3}\gamma_t^\alpha\gamma_t^\mu)\gamma_t^\nu \gamma_5 g_s  {\lambda_{eb}^m\over2}G_{\mu\nu}^m h_v^c\, ,
\\ \nonumber  
\widetilde{J}_{6_f,6_c,0,1,3/2^-}^\alpha &=& \epsilon_{aec}[q^{aT}\mathcal{C}\gamma_5 q^{b}+q^{bT}\mathcal{C}\gamma_5 q^{a}]
\\  &\times& (g_t^{\alpha\mu}-{1\over3}\gamma_t^\alpha\gamma_t^\mu)\gamma_t^\nu \gamma_5 g_s  {\lambda_{eb}^m\over2}\widetilde{G}_{\mu\nu}^m h_v^c\, .
\end{eqnarray}

\item The singlet $[\bar{3}_f,6_c,1,0,G]$ contains one state with $J^P=1/2^+$, and its dual singlet $[\bar{3}_f,6_c,1,0,\widetilde{G}]$ contains another state with $J^P=1/2^-$. Their corresponding currents are
\begin{eqnarray}
\nonumber J_{\bar{3}_f,6_c,1,0,1/2^+} &=& \epsilon_{abe}[q^{aT}\mathcal{C}\sigma_t^{\mu\nu} \gamma_5 q^{b}+q^{bT}\mathcal{C}\sigma_t^{\mu\nu} \gamma_5 q^{a}]
\\ && ~~~~~~~~~~~~~~~~~~~~~~~~~~~~~~~ \times~~ g_s {\lambda_{ec}^m\over2 }G_{\mu\nu}^m h_v^c\, ,
\\ \nonumber
\widetilde{J}_{\bar{3}_f,6_c,1,0,1/2^-} &=& \epsilon_{abe}[q^{aT}\mathcal{C}\sigma_t^{\mu\nu} \gamma_5 q^{b}+q^{bT}\mathcal{C}\sigma_t^{\mu\nu} \gamma_5 q^{a}]
\\ && ~~~~~~~~~~~~~~~~~~~~~~~~~~~~~~~ \times~~ g_s {\lambda_{ec}^m\over2}\widetilde{G}_{\mu\nu}^m h_v^c\, .
\end{eqnarray}

\item The doublet $[\bar{3}_f,6_c,1,1,G]$ contains two states with $J^P=1/2^+$ and $3/2^+$. Its dual doublet $[\bar{3}_f,6_c,1,1,\widetilde{G}]$ contains two other states with $J^P=1/2^-$ and $3/2^-$. Their corresponding currents are
\begin{eqnarray}
\nonumber J_{\bar{3}_f,6_c,1,1,1/2^+} &=& \epsilon_{abe}[q^{aT}\mathcal{C}\gamma_t^\mu q^{b}+q^{bT}\mathcal{C}\gamma_t^\mu q^{a}]
\\ && ~~~~~~~~~~~\times~~ \gamma_t^{\nu} \gamma_5 g_s {\lambda_{ec}^m\over 2}G_{\mu\nu}^m h_v^c\, ,
\\ \nonumber
\widetilde{J}_{\bar{3}_f,6_c,1,1,1/2^-} &=& \epsilon_{abe}[q^{aT}\mathcal{C}\gamma_t^\mu q^{b}+q^{bT}\mathcal{C}\gamma_t^\mu q^{a}]
\\ && ~~~~~~~~~~~\times~~ \gamma_t^{\nu} \gamma_5 g_s {\lambda_{ec}^m\over2}\widetilde{G}_{\mu\nu}^m h_v^c\, ,
\\ \nonumber
J_{\bar{3}_f,6_c,1,1,3/2^+}^{\alpha} &=& \epsilon_{abe}[q^{aT}\mathcal{C}\gamma_t^\mu q^{b}+q^{bT}\mathcal{C}\gamma_t^\mu q^{a}]
\\ &\times& (g^{\alpha\nu}_t-{1\over3}\gamma_t^{\alpha}\gamma_t^\nu)g_s {\lambda_{ec}^m\over2}G_{\mu\nu}^m h_v^c\, ,
\\ \nonumber
\widetilde{J}_{\bar{3}_f,6_c,1,1,3/2^-}^{\alpha} &=& \epsilon_{abe}[q^{aT}\mathcal{C}\gamma_t^\mu q^{b}+q^{bT}\mathcal{C}\gamma_t^\mu q^{a}]
\\ &\times& (g^{\alpha\nu}_t-{1\over3}\gamma_t^{\alpha}\gamma_t^\nu) g_s {\lambda_{ec}^m\over2}\widetilde{G}_{\mu\nu}^m h_v^c\, .
\end{eqnarray}

\item The doublet $[\bar{3}_f,6_c,1,2,G]$ contains two states with $J^P=3/2^+$ and $5/2^+$. Its dual doublet $[\bar{3}_f,6_c,1,2,\widetilde{G}]$ contains two other states with $J^P=3/2^-$ and $5/2^-$. Their corresponding currents are
\begin{eqnarray}
\nonumber J_{\bar{3}_f,6_c,1,2,3/2^+}^{\alpha} &=& \epsilon_{abe}[q^{aT}\mathcal{C}\gamma_{t,\mu_1} q^{b}+q^{bT}\mathcal{C}\gamma_{t,\mu_1} q^{a}]
\\ && ~ \times~~ \Gamma_{J=2}^{\alpha\mu_4,\mu_1\nu_2}\gamma_t^{\mu_4}\gamma_t^{\nu_1} g_s {\lambda_{ec}^m\over2}G_{\nu_1\nu_2}^m h_v^c\, ,
\\ \nonumber
\widetilde{J}_{\bar{3}_f,6_c,1,2,3/2^-}^{\alpha} &=& \epsilon_{abe}[q^{aT}\mathcal{C}\gamma_{t,\mu_1} q^{b}+q^{bT}\mathcal{C}\gamma_{t,\mu_1} q^{a}] 
\\ && ~ \times~~ \Gamma_{J=2}^{\alpha\mu_4,\mu_1\nu_2}\gamma_t^{\mu_4}\gamma_t^{\nu_1} g_s {\lambda_{ec}^m\over2}\widetilde{G}_{\nu_1\nu_2}^m h_v^c\, ,
\\ \nonumber 
J_{\bar{3}_f,6_c,1,2,5/2+}^{\alpha\beta} &=& \epsilon_{abe}[q^{aT}\mathcal{C}\gamma_{t,\mu_1} q^{b}+q^{bT}\mathcal{C}\gamma_{t,\mu_1} q^{a}]
\\ && ~~~~ \times~~ \Gamma_{J={5\over 2}}^{\alpha\beta,\mu_1\nu_2}\gamma_t^{\nu_1}\gamma_5 g_s {\lambda_{ec}^m\over2}G_{\nu_1\nu_2}^m h_v^c\, ,
\\ \nonumber
\widetilde{J}_{\bar{3}_f,6_c,1,2,5/2-}^{\alpha\beta} &=& \epsilon_{abe}[q^{aT}\mathcal{C}\gamma_{t,\mu_1} q^{b}+q^{bT}\mathcal{C}\gamma_{t,\mu_1} q^{a}]
\\ && ~~~~ \times~~ \Gamma_{J={5\over 2}}^{\alpha\beta,\mu_1\nu_2}\gamma_t^{\nu_1}\gamma_5 g_s {\lambda_{ec}^m\over2}\widetilde{G}_{\nu_1\nu_2}^m h_v^c\, .
\label{eq:end1}
\end{eqnarray}

\end{itemize}
In the above expressions, $\gamma_t^\mu=\gamma^\mu-v\!\!\!\slash v^\mu$, $g_t^{\mu\nu}=g^{\mu\nu}-v^{\mu}v^{\nu}$, $\sigma_t^{\mu\nu}=\sigma^{\mu\nu}-\sigma^{\mu\alpha}v_\alpha v^\nu-\sigma^{\alpha\nu}v_\alpha v^\mu$, and $G_{\mu\nu}^n=\partial_\mu A_\nu^n-\partial_\nu A_\mu^n +g_s f^{npq} A_{p,\mu}A_{q,\nu}$; $\Gamma_{J=2}^{\alpha\beta,\mu\nu}$ and $\Gamma_{J=5/2}^{\alpha\beta,\mu\nu}$ are the $J=2$ and $J=5/2$ projection operators:
\begin{eqnarray}
\Gamma_{J=2}^{\alpha\beta,\mu\nu}&=&g_t^{\alpha\mu}g_t^{\beta\nu}+g_t^{\alpha\nu}g_t^{\beta\mu}-{2\over3}g_t^{\alpha\beta}g_t^{\mu\nu} \, ,
\\
\Gamma_{J=5/2}^{\alpha\beta,\mu\nu}&=&g_t^{\alpha\mu}g_t^{\beta\nu}+g_t^{\alpha\nu}g_t^{\beta\mu}-{2\over5}g_t^{\alpha\beta}g_t^{\mu\nu}
\\ \nonumber &-& {1\over5}g_t^{\alpha\mu}\gamma_t^{\beta}\gamma_t^\nu
- {1\over5}g_t^{\alpha\nu}\gamma_t^\beta\gamma_t^\mu
- {1\over5}g_t^{\beta\mu}\gamma_t^\alpha\gamma_t^\nu
- {1\over5}g_t^{\beta\nu}\gamma_t^\alpha\gamma_t^\mu\, .
\end{eqnarray}

\section{Results at the leading order}
\label{sec:leading}

In this section we apply the QCD sum rule method to investigate the charmed hybrid baryon currents given in Eqs.~(\ref{eq:start}-\ref{eq:end}) within the framework of heavy quark effective theory. These seven currents form the four baryon multiplets $[\bar{3}_f,\bar{3}_c,0,1,G]$, $[6_f,\bar{3}_c,1,0,G]$, $[6_f,\bar{3}_c,1,1,G]$, and $[6_f,\bar{3}_c,1,2,G]$. The other currents given in Eqs.~(\ref{eq:start1}-\ref{eq:end1}) will be investigated in our future studies. We shall find that the current $J_{6_f,\bar{3}_c,1,1,1/2^+}$, belonging to the doublet $[6_f,\bar{3}_c,1,1,G]$, couples to the lowest-lying charmed hybrid baryon, so we use this current as a representative example. The explicit forms of this current and its spin partner for the $\Sigma_{cg}$ states are presented as follows ($q=up/down$):
\begin{eqnarray}
&& J_{\Sigma_{cg},\bar{3}_c,1,1,1/2^+}=\epsilon_{abe}[q^{aT}\mathcal{C}\gamma_t^\mu q^{b}]\gamma_t^{\nu} \gamma_5 g_s {\lambda_{ec}^m\over2}G_{\mu\nu}^m h_v^c\, ,
\label{eq:12}
\\ \nonumber
&& J_{\Sigma_{cg},\bar{3}_c,1,1,3/2^+}^{\alpha}=\epsilon_{abe}[q^{aT}\mathcal{C}\gamma_t^\mu q^{b}](g^{\alpha\nu}_t-{1\over3}\gamma_t^{\alpha}\gamma_t^\nu) 
g_s {\lambda_{ec}^m\over2}G_{\mu\nu}^m h_v^c \, .
\\ \label{eq:32}
\end{eqnarray}

The current $J_{\Sigma_{cg},\bar{3}_c,1,1,1/2^+}$ couples to both the charmed hybrid baryon with $J^P=1/2^+$ and the one with $J^P=1/2^-$ through
\begin{eqnarray}
\langle 0|J_{\Sigma_{cg},\bar{3}_c,1,1,1/2^+}|\Sigma_{cg}^+\rangle &=& f_+ u \, ,
\\ \langle 0|J_{\Sigma_{cg},\bar{3}_c,1,1,1/2^+}|\Sigma_{cg}^-\rangle &=& f_- \gamma_5 u \, ,
\end{eqnarray}
where $f_\pm$ are the decay constants and $u$ is the Dirac spinor. Consequently, the two-point correlation function can be constructed as 
\begin{eqnarray}
\nonumber &&\Pi_{\Sigma_{cg},\bar{3}_c,1,1,1/2^+}(\omega)
\\
\nonumber &=& i\int d^4 x e^{ikx}\langle 0|T [J_{\Sigma_{cg},\bar{3}_c,1,1,1/2^+}(x)\bar{J}_{\Sigma_{cg},\bar{3}_c,1,1,1/2^+}(0)]|0\rangle
\\ &=& {1+v\!\!\!\slash\over2}\Pi_+(\omega)
 + {1-v\!\!\!\slash\over2}\Pi_-(\omega) \, ,
\label{eq:two-point}
\end{eqnarray}
where $\omega=v\cdot k$ is the external off-shell energy, and $\Pi_\pm$ are respectively contributed by $\Sigma_{cg}^\pm$. We calculate Eq.~(\ref{eq:two-point}) at the quark-gluon level using the operator product expansion (OPE) method to derive
\begin{equation}
\Pi_-(\omega) = 0 \, ,
\label{eq:negative}
\end{equation}
suggesting that the current $J_{\Sigma_{cg},\bar{3}_c,1,1/2^+}$ fully couples to the positive-parity state $\Sigma_{cg}^+$. This allows us to simplify Eq.~(\ref{eq:two-point}) at the hadron level as
\begin{equation}
\label{eq:pole}
\Pi_+(\omega) = { f^2_+ \over \bar{\Lambda}_+ - \omega} + \rm{higher~~states} \, ,
\end{equation}
with $\bar{\Lambda}_+$ the sum rule result at the leading order. After performing the Borel transformation at both the hadron and quark-gluon levels, we obtain
\begin{eqnarray}
\nonumber 
\Pi_+(\omega_c, T) &=& f_+^2 e^{-\bar{\Lambda}_+/T}
\\ \nonumber &=& \int_0^{\omega_c}e^{-\omega/T}d\omega \times \Big({\alpha_s \over 1890 \pi^5}\omega^9
+{\langle g_s^2 GG\rangle\over 240 \pi^4}\omega^9
\\ \nonumber && -{\alpha_s \langle g_s^2 GG \rangle\over 1440 \pi^5}\omega^5
-{\langle g_s^3 G^3\rangle\over 192 \pi^4}\omega^3 
+{8 \alpha_s \langle \bar q q \rangle^2\over 9 \pi}\omega^3
\\ \nonumber && +{\langle g_s^2 G G\rangle^2\over 4608 \pi^4}\omega
-{41\alpha_s \langle\bar q q\rangle\langle g_s \bar q \sigma G q \rangle\over 48 \pi}\omega\Big)
\\ \nonumber && +{\langle \bar q q \rangle^2\langle g_s^2 GG\rangle \over 72}-{\langle\bar q q\rangle\langle g_s^2 GG\rangle\langle g_s \bar q \sigma G q \rangle\over 576 T^2}
\\ && -{\langle \bar q q \rangle^2 \langle g_s^3 G^3\rangle \over 576 T^2}+{25\alpha_s \langle \bar q \sigma G q\rangle^2\over 768 \pi}\, ,
\label{eq:positive}
\end{eqnarray}
from which we further derive
\begin{eqnarray}
\overline{\Lambda}_+(\omega_c, T) &=& \frac{1}{\Pi_+(\omega_c, T)} \times \frac{\partial \Pi_+(\omega_c, T)}{\partial(-1/T)} \, ,
\label{eq:mass}
\\
f_+^2(\omega_c, T) &=& \Pi_+(\omega_c, T) \times e^{\overline{\Lambda}_+(\omega_c, T) / T} \, .
\label{eq:coupling}
\end{eqnarray}
In the above expressions, we have evaluated the two-point correlation function up to the twelfth dimension, including the quark condensate $\langle \bar q q\rangle$, the two-gluon condensate $\langle g_s^2 GG\rangle$, the three-gluon condensate $\langle g_s^3 G^3\rangle$, the quark-gluon mixed condensate $\langle g_s \bar q \sigma G q \rangle$, and their combinations. The sum rule equations for other currents belonging to the multiplets $[\bar{3}_f,\bar{3}_c,0,1,G]$, $[6_f,\bar{3}_c,1,0,G]$, $[6_f,\bar{3}_c,1,1,G]$, and $[6_f,\bar{3}_c,1,2,G]$ are provided in the supplemental Mathematica file ``OPE.nb''.

To perform numerical analyses, we use the following values for various quark and gluon condensate~\cite{ParticleDataGroup:2020ssz,Ovchinnikov:1988gk,Yang:1993bp,Ellis:1996xc,Ioffe:2002be,Jamin:2002ev,Gimenez:2005nt,Narison:2011xe,Narison:2018dcr}:
\begin{eqnarray}
\nonumber \langle \bar q q\rangle &=&-(0.240\pm 0.010)^3~\rm{GeV}\, ,
\\ \nonumber \langle \bar s s \rangle &=& (0.8\pm 0.1)\times \langle \bar q q\rangle\, ,
\\ \langle \alpha_s GG\rangle &=&(6.35\pm 0.35)\times 10^{-2} \rm{GeV}^4\, ,
\label{eq:condensate}
\\ \nonumber \langle g_s^3 G^3 \rangle &=& (8.2\pm 1.0)\times \langle \alpha_s GG\rangle ~\rm{GeV}^2\, ,
\\ \nonumber \langle g_s \bar q \sigma G q \rangle &=& (0.8\pm 0.2)\times \langle \bar q q\rangle ~\rm{ GeV}^2 \, ,
\\ \nonumber \langle g_s \bar s \sigma G s \rangle &=& (0.8\pm 0.2)\times \langle \bar s s\rangle~\rm{ GeV}^2 \, .
\end{eqnarray}
Additionally, we use the following values for the running masses of the charm and strange quarks at the renormalization scale of $2$~GeV~\cite{Chen:2013zia}:
\begin{eqnarray}
\nonumber \alpha_s(\mu=M_\tau)&=&0.33\, ,
\\ \nonumber \alpha_s(\mu=2~{\rm GeV})&=& {\alpha_s(M_\tau)\over 1+{25 \alpha_s(M_\tau)\over 12 \pi}\log({\mu^2\over M_\tau^2})} = 0.31 \, , 
\\ \overline{m}_c(\mu=m_c) &=& {1.23\pm 0.09}~\rm{GeV}\, ,
\label{eq:quarkmass}
\\ \nonumber m_c(\mu=2~{\rm GeV})&=&\overline{m}_c({\alpha_s(\mu)\over \alpha_s(\overline{m}_c)} )^{12/25} = 1.10~{\rm GeV}\, ,
\\ \nonumber m_s(\mu=2~{\rm GeV}) &=& 93^{+11}_{-~5}~\rm{MeV}\, .
\end{eqnarray} 

The sum rule equation listed in Eq.~(\ref{eq:positive}) contains two free parameters: the threshold value $\omega_c$ and the Borel mass $T$. Their appropriate working regions are determined based on three criteria: a) sufficient convergence of the OPE, b) a sufficiently large pole contribution, and c) weak dependence of the sum rule results on these parameters. 

\begin{figure}[H]
\centering
\scalebox{0.8}{\includegraphics{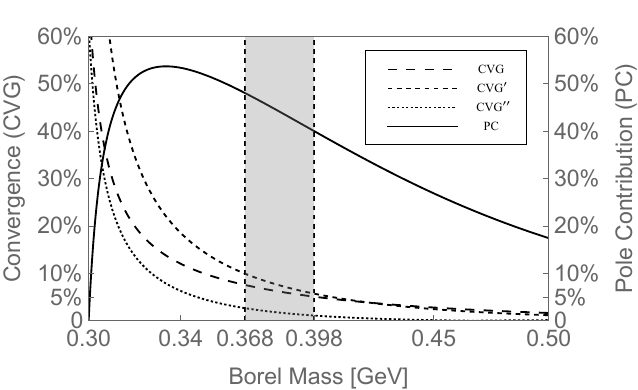}}
\caption{CVG$^{(\prime,\prime\prime)}$ and PC as functions of the Borel mass $T$, with the threshold value set to be $\omega_c = 2.95$~GeV. These curves are derived using the current $J_{\Sigma_{cg}^+,\bar 3_c,1,1,1/2^+}$, which belongs to the doublet $[\Sigma_{cg}^+,\bar 3_c,1,1, G]$.} 
\label{fig:condition}
\end{figure}

To ensure sufficient convergence of the OPE, we require the $g_s^4$ terms to be less than $5\%$, the $D=8$ terms to be less than $10\%$, and the $D=6$ terms to be less than $20\%$:
\begin{eqnarray}
\nonumber \rm{CVG}&\equiv&|{\Pi^{g_s^n=4}_+(\infty, T)\over \Pi_+(\infty,T)}|\leq 5\%\, ,
\\ \rm{ CVG^\prime}&\equiv&|{\Pi^{D=8}_+(\infty, T)\over \Pi_+(\infty,T)}|\leq 10\%\, ,
\\ \nonumber \rm{CVG^{\prime\prime}}&\equiv&|{\Pi^{D=6}_+(\infty, T)\over \Pi_+(\infty,T)}|\leq 20\%\, .
\end{eqnarray}
As shown in Fig.~\ref{fig:condition}, the lower limit of the Borel mass is determined to be $T_{min} = 0.37$~GeV. 

To ensure a sufficiently large pole contribution, we require the pole contribution to exceed $40\%$:
\begin{eqnarray}
\rm{PC}&\equiv& |{\Pi_+(\omega_0,T)\over \Pi_+(\infty,T)}|\ge 40\%\, .
\end{eqnarray}
As shown in Fig.~\ref{fig:condition}, the upper limit of the Borel mass is determined to be $T_{max} = 0.40$~GeV, when setting $\omega_c= 2.95~\rm{GeV}$. 

\begin{figure*}[hbt]
\begin{center}
\subfigure[]{
\scalebox{0.7}{\includegraphics{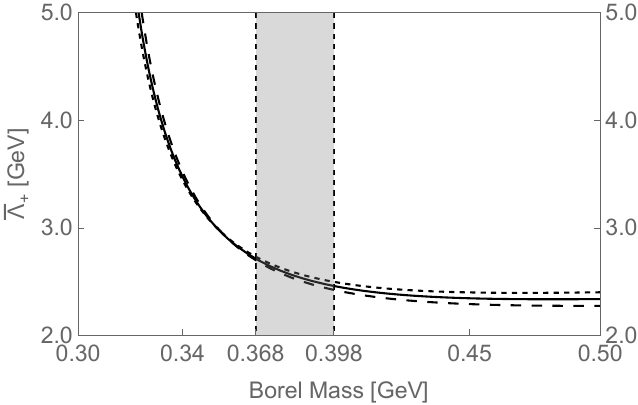}}}~~~~
\subfigure[]{
\scalebox{0.73}{\includegraphics{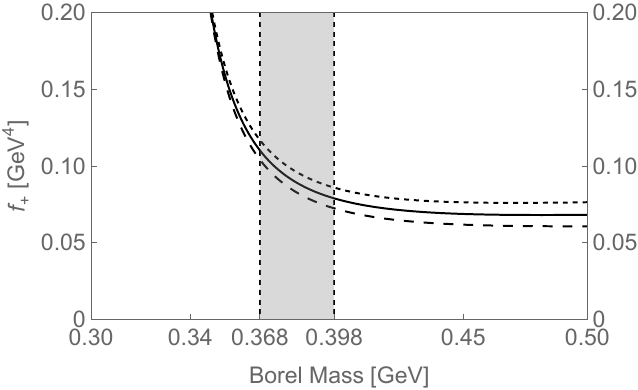}}}
\caption{Variations of (a) $\bar{\Lambda}_+$ and (b) $f_+$ as functions of the Borel mass $T$, where the long-dashed, solid, and short-dashed curves correspond to fixed threshold values $\omega_c = 2.85~\mathrm{GeV}$, $2.95~\mathrm{GeV}$, and $3.05~\mathrm{GeV}$, respectively. These results are obtained using the current $J_{\Sigma_{cg}^+,\bar{3}c,1,1,1/2^+}$, which is associated with the doublet $[\Sigma_{cg}^+,\bar{3}_c,1,1,G]$.}
\label{fig:leading}
\end{center}
\end{figure*}

Altogether, we obtain the Borel window $0.37~\rm{GeV}\leq T\leq 0.40~\rm{GeV}$ for $\omega_c= 2.95~\rm{GeV}$. By varying $\omega_c$, we find that non-vanishing Borel windows exist for $\omega_c\ge \omega_c^{min}=2.7~\rm{GeV}$. We choose $\omega_c$ to be approximately $10\%$ larger and establish its working region as $2.7~ \rm{GeV}\leq \omega_c \leq 3.2~\rm{GeV}$, where the numerical results are derived as:
\begin{eqnarray}
\bar{\Lambda}_+ &=& 2.46^{+0.15}_{-0.14}~\rm{GeV} \, ,
\\ f_+ &=& 0.079^{+0.015}_{-0.013}~\rm{GeV}^4 \, .
\end{eqnarray}
Their variations are shown in Fig.~\ref{fig:leading} as functions of the Borel mass $T$, where the dependence is weak and remains within an acceptable range inside the Borel window $0.37~\rm{GeV} \leq T \leq 0.40~\rm{GeV}$.

\section{Results at the ${\mathcal O}(1/m_Q)$ order}
\label{sec:nexttoleading}

In this section we extend the analysis to include ${\mathcal O}(1/m_Q)$ corrections by considering the following Lagrangian from the heavy quark effective theory
\begin{eqnarray}
\mathcal{L}_{\rm eff} = \bar{h}_{v}^civ\cdot D_t h_{v}^c + \frac{1}{2m_c}\mathcal{K} + \frac{1}{2m_c}\mathcal{S} \, ,
\label{eq:next}
\end{eqnarray}
where $\mathcal{K}$ represents the operator of nonrelativistic kinetic energy and $\mathcal S$ represents the Pauli term describing the chromomagnetic interaction:
\begin{eqnarray}
\mathcal{K} &=& \bar{h}_{v}^c(iD_{t})^{2}h_{v}^c \, ,
\\
\mathcal{S} &=& \frac{g_s}{2} C_{mag} (m_c/\mu) \bar{h}_{v}^c \sigma_{\mu\nu} G^{\mu\nu} h_{v}^c \, ,
\label{eq:Sdefine}
\end{eqnarray}
with $C_{mag} (m_{c}/\mu) = [ \alpha_s(m_c) / \alpha_s(\mu) ]^{3/\beta_0}$ and $\beta_0 = 11 - 2 n_f /3$.

Following Eq.~(\ref{eq:pole}), we can express the pole term up to the ${\mathcal O}(1/m_c)$ order as
\begin{eqnarray}
\nonumber \Pi_+(\omega) &=& \frac{(f_+ + \delta f_+ )^{2}}{\overline{\Lambda}_+ + \delta M_+ - \omega}
\\ &=& \frac{f_+^{2}}{\overline{\Lambda}_+-\omega}-\frac{\delta M_+ f_+^{2}}{(\overline{\Lambda}_+-\omega)^{2}}+\frac{f_+ \delta f_+}{\overline{\Lambda}_+-\omega} \, ,
\label{eq:correction}
\end{eqnarray}
where $\delta M_+$ and $\delta f_+$ denote the corrections to the mass and decay constant, respectively. To calculate $\delta M_+$, we consider the following three-point correlation functions
\begin{eqnarray}
\nonumber && \delta_{\mathcal{O}}\Pi_{F,C,j_{qq},j_l,J^P}^{\alpha_{1}\cdots\alpha_{J-1/2},\beta_{1}\cdots\beta_{J-1/2}}(\omega, \omega^\prime)
\\ \nonumber &=& i^{2}\int d^{4}xd^{4}y e^{i k x - i k^\prime y}\langle0|TJ_{F,C,j_{qq},j_l,J^P}^{\alpha_{1}\cdots \alpha_{J-1/2}}(x) \mathcal{O} \bar J_{F,C,j_{qq},j_l,J^P}^{\beta_{1}\cdots \beta_{J-1/2}}(y)|0\rangle
\\ \nonumber &=& \mathbb{S} [ g_t^{\alpha_1 \beta_1} \cdots g_t^{\alpha_{J-1/2} \beta_{J-1/2}} ]
\\ &\times& \left( {1+v\!\!\!\slash\over2}\delta_{\mathcal{O}} \Pi^+_{F,C,j_{qq},j_l,J^P} + {1-v\!\!\!\slash\over2}\delta_{\mathcal{O}} \Pi^-_{F,C,j_{qq},j_l,J^P} \right) \, ,
\label{eq:nextpi}
\end{eqnarray}
where $\mathcal{O} = \mathcal{K}$ or $\mathcal{S}$, $k^\prime = k + q$, $\omega = v \cdot k$, and $\omega^\prime = v \cdot k^\prime$. 

Based on Eq.~(\ref{eq:next}), we can express Eq.~(\ref{eq:nextpi}) at the hadron level as
\begin{eqnarray}
\delta_{\mathcal{K}} \Pi_+
&=& \frac{f_+^{2}K_+}{(\overline{\Lambda}_+-\omega)(\overline{\Lambda}_+-\omega^\prime)}
+\frac{f_+^{2}G_{\mathcal{K}}}{\overline{\Lambda}_+-\omega}
+\frac{f_+^{2}G^\prime_{\mathcal{K}}}{\overline{\Lambda}_+-\omega^\prime} \, ,
\\ \nonumber \delta_{\mathcal{S}}\Pi_+
&=& \frac{d_{M}f_+^{2}\Sigma_+}{(\overline{\Lambda}_+-\omega)(\overline{\Lambda}_+-\omega^\prime)}
+\frac{d_{M}f_+^{2}G_{\mathcal{S}}}{\overline{\Lambda}_+-\omega} \,
+\frac{d_{M}f_+^{2}G^\prime_{\mathcal{S}}}{\overline{\Lambda}_+-\omega^\prime} \, ,
\\
\end{eqnarray}
where $K_+$, $\Sigma_+$, and $d_{M}$ are defined as
\begin{eqnarray}
\nonumber K_+ &\equiv& \langle \Sigma_{cg}^+ |\bar{h}_{v}^c(iD_t)^{2}h_{v}^c| \Sigma_{cg}^+ \rangle \, ,
\\ \nonumber d_{M}\Sigma_+ &\equiv& \langle \Sigma_{cg}^+ | {g_s\over2} \bar{h}_{v}^c\sigma_{\mu\nu}G^{\mu\nu}h_{v}^c| \Sigma_{cg}^+ \rangle \, ,
\\ d_{M} &\equiv& d_{j,j_{l}} \, ,
\\ \nonumber d_{j_{l}-1/2,j_{l}} &=& 2j_{l}+2\, ,
\\ \nonumber d_{j_{l}+1/2,j_{l}} &=& -2j_{l} \, .
\end{eqnarray}
Note that the term $\mathcal S$ causes a mass splitting within the same doublet, while it can also lead to a mixing of states with the same spin-parity quantum number. We have accounted for the former effect, but the latter effect is found in Ref.~\cite{Dai:1998ve} to be negligible and is therefore not considered in the present study.

\begin{figure*}[hbt]
\begin{center}
\subfigure[]{
\scalebox{0.735}{\includegraphics{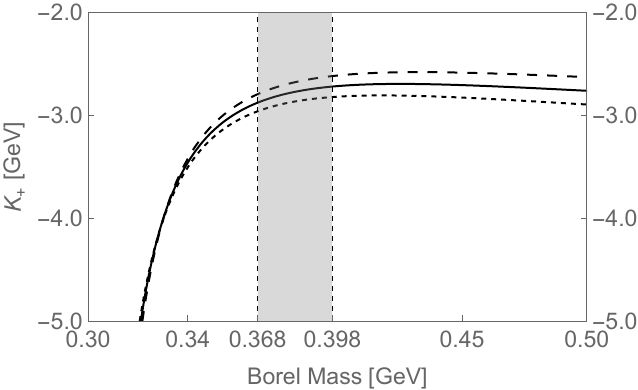}}}~~~~
\subfigure[]{
\scalebox{0.7}{\includegraphics{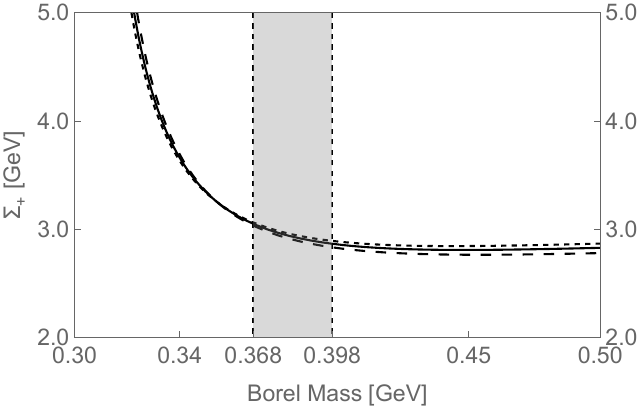}}}
\caption{Variations of (a) $K_+$ and (b) $\Sigma_+$ as functions of the Borel mass $T$, where the long-dashed, solid, and short-dashed curves correspond to fixed threshold values $\omega_c = 2.85~\mathrm{GeV}$, $2.95~\mathrm{GeV}$, and $3.05~\mathrm{GeV}$, respectively. These results are obtained using the current $J_{\Sigma_{cg}^+,\bar{3}c,1,1,1/2^+}$, which is associated with the doublet $[\Sigma_{cg}^+,\bar{3}_c,1,1,G]$.}
\label{fig:KS}
\end{center}
\end{figure*}

\begin{figure*}[hbt]
\begin{center}
\subfigure[]{
\scalebox{0.7}{\includegraphics{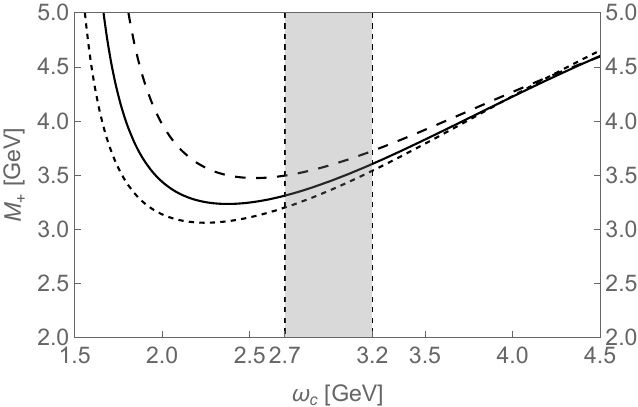}}}
~~~~
\subfigure[]{
\scalebox{0.7}{\includegraphics{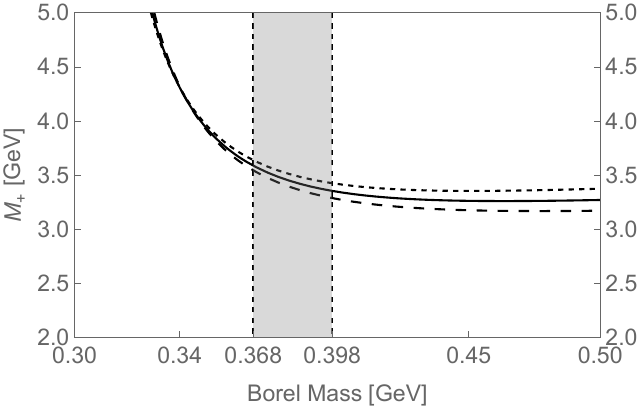}}}
\caption{Variations of $M_+$ as functions of (a) the threshold value $\omega_c$ and (b) the Borel mass $T$. In the left panel the long-dashed, solid, and short-dashed curves correspond to fixed Borel masses $T = 0.37~\mathrm{GeV}$, $0.38~\mathrm{GeV}$, and $0.40~\mathrm{GeV}$, respectively. In the right panel the long-dashed, solid, and short-dashed curves correspond to fixed threshold values $\omega_c = 2.85~\mathrm{GeV}$, $2.95~\mathrm{GeV}$, and $3.05~\mathrm{GeV}$, respectively. These results are obtained using the current $J_{\Sigma_{cg}^+,\bar{3}c,1,1,1/2^+}$, which is associated with the doublet $[\Sigma_{cg}^+,\bar{3}_c,1,1,G]$.}
\label{fig:mass}
\end{center}
\end{figure*}

The correlation functions in Eq.~(\ref{eq:nextpi}) can also be evaluated at the quark-gluon level using the operator product expansion. After performing the double Borel transformation to convert $\omega$ and $\omega^\prime$ into $T_1$ and $T_2$, and setting these two Borel parameters equal, we arrive at
\begin{eqnarray}
\nonumber && \delta_{\mathcal{K}}\Pi_+ (\omega_c,T) = f_+^2 K_+ e^{-\bar{\Lambda}_+/T}
\\ \nonumber &=& \int_0^{\omega_c} e^{-\omega/ T} d\omega \times \Big(
-{\alpha_s\over 4950 \pi^5}\omega^{11}
-{11\langle g_s^2 GG\rangle\over 5040 \pi^4}\omega^7
\\ \nonumber && 
+{13\alpha_s \langle g_s^2 GG \rangle\over 24192 \pi^5}\omega^7
+{\langle g_s^3 G^3\rangle\over 1920 \pi^4}\omega^5
-{8\alpha_s \langle \bar q q\rangle^2\over 9\pi}\omega^5
\\ \nonumber && 
+{\langle g_s^2 GG\rangle^2\over 256 \pi^4}\omega^3-{47\langle g_s^2 GG\rangle^2\over 110592\pi^4}\omega^3
\\ \nonumber &~~~~&+{269 \alpha_s\langle \bar q q\rangle\langle g_s \bar q\sigma G q\rangle\over 144\pi}\omega^3-{41\alpha_s \langle g_s \bar q \sigma G q\rangle^2\over 192\pi}\omega\Big)
\\ && 
-{5\langle\bar q q\rangle\langle g_s^2GG\rangle\langle g_s\bar q \sigma G q\rangle\over 512}-{\langle\bar q q\rangle^2\langle g_s^3 G^3\rangle\over 288}\, ,
\label{eq:K}
\\ 
\nonumber && \delta_{\mathcal{S}}\Pi_+ (\omega_c,T) = 4 \times f_+^2 \Sigma_+ e^{-\bar{\Lambda}_+/T}
\\ \nonumber &=& \int_0^{\omega_c} e^{-\omega/ T} d\omega \times \Big(
{815\alpha_s \langle g_s^2 GG\rangle\over 24192\pi^5}\omega^7
+{\langle g_s^3 G^3\rangle\over 960\pi^4}\omega^5
\\ \nonumber && 
+{251\langle g_s^2 GG\rangle^2\over 221184\pi^4}\omega^3
-{511\alpha_s\langle\bar q q\rangle\langle g_s\bar q\sigma G q\rangle\over108\pi}\omega^3
\\ \nonumber && 
+{8779\alpha_s\langle g_s\bar q \sigma G q\rangle^2\over 4608\pi}\omega
+{9\alpha_s\langle\bar q q^2\rangle\langle g_s^2 GG\rangle\over 32\pi}\omega\Big)
\\ \nonumber && 
-{31\langle\bar q q\rangle\langle g_s^2 GG\rangle\langle g_s \bar q \sigma G q \rangle\over 6912}
+{\langle\bar q q\rangle^2\langle g_s^3 G^3\rangle\over 288}
\\ && 
-{9\alpha_s\langle \bar q q\rangle\langle g_s^2 GG\rangle \langle g_s\bar q\sigma G q\rangle\over 256\pi}\, .
\label{eq:S} 
\end{eqnarray}
The Borel mass dependence of these sum rule equations is shown in Fig.~\ref{fig:KS}. Both $K$ and $\Sigma$ exhibit mild variation with $T$ in the working region $2.7~ \rm{GeV}\leq \omega_c \leq 3.2~\rm{GeV}$. From this region, we extract the following numerical results:
\begin{eqnarray}
K_+ &=& -2.72^{+0.19}_{-0.20}~\rm{GeV}^2 \, ,
\label{eq:Kvalue}
\\ \Sigma_+ &=& 2.87\pm 0.25~\rm{GeV}^2 \, .
\label{eq:Svalue}
\end{eqnarray}
These two values represent the sum rule results at the $\mathcal{O}(1/m_c)$ order, from which we derive
\begin{eqnarray}
\delta M_+ &=& -\frac{1}{2m_c}\left(K_+ + d_{M}C_{mag}\Sigma_+ \right) \, ,
\end{eqnarray}
and further extract the mass of the $\Sigma_{cg}^+$ state as
\begin{eqnarray}
M_{\Sigma_{cg}(1/2^+)} &\equiv& M_+ = m_c + \overline{\Lambda}_+ + \delta M_+ \\
\nonumber &=& 1.10^{+0.08}_{-0.08}~\mathrm{GeV} + 2.46^{+0.15}_{-0.14}~\mathrm{GeV} - 0.21^{+0.13}_{-0.12}~\mathrm{MeV} \\
\nonumber &=& 3.36^{+0.27}_{-0.26}~\mathrm{GeV} \, ,
\end{eqnarray}
whose dependence on the threshold value $\omega_c$ and the Borel mass $T$ is illustrated in Fig.~\ref{fig:mass}. The uncertainties originate from variations in the threshold value $\omega_c$, the Borel mass $T$, and the QCD parameters provided in Eqs.~(\ref{eq:condensate}) and (\ref{eq:quarkmass}). 

In addition, the mass splitting between the $\Sigma_{cg}$ states of $J^P = 1/2^+$ and $J^P = 3/2^+$ is derived as
\begin{eqnarray}
\Delta M_+ = M_{\Sigma_{cg}(3/2^+)} - M_{\Sigma_{cg}(1/2^+)} = 2.16^{+0.32}_{-0.29}~\mathrm{GeV} \, .
\end{eqnarray}
It is important to note that this mass splitting is rather large, an effect that should vanish in the heavy quark limit. As seen from Eq.~(\ref{eq:Sdefine}), the term $\mathcal{S}$ is directly related to the mass splitting, and this term is significantly influenced by the gluon degrees of freedom. Therefore, the presence of a significant gluonic component in the charmed hybrid baryon might result in a substantial mass difference. Furthermore, as shown in Eq.~(\ref{eq:Kvalue}) and Eq.~(\ref{eq:Svalue}), both the $K$ and $\Sigma$ terms are ${\mathcal O}(1/m_Q)$ corrections, and their magnitudes are of the same order, indicating that the observed mass splitting is consistent with the ${\mathcal O}(1/m_Q)$ correction scale. The effects of these gluonic contributions will be addressed further in future studies to quantify their impact more rigorously.

Similarly, we apply the QCD sum rule method to study the other currents belonging to the multiplets $[\bar{3}_f,\bar{3}_c,0,1,G]$, $[6_f,\bar{3}_c,1,1,G]$, $[6_f,\bar{3}_c,1,2,G]$, and $[6_f,\bar{3}_c,1,0,G]$. The obtained results are summarized in Table~\ref{tab:results}, where the current $J_{6_f,\bar{3}_c,1,1,1/2^+}$ belonging to the doublet $[6_f,\bar{3}_c,1,1,G]$ couples to the lowest-lying charmed hybrid baryon. A point of confusion arises because this current exhibits a symmetric flavor structure for the two light quarks, which contrasts with the flavor structure of the lowest-lying charmed baryon $\Lambda_{cg}$, but is consistent with the flavor structure of $\Sigma_{cg}$. To explain the possible reasons, we notice that there are two states with the flavor structure $\bar{3}_f$, but five with $6_f$. Hence, if the mass of one of the five states with $6_f$ were to decrease, it would become possible for this state to become the lowest-lying one, leading to the situation where the lowest-lying charmed baryons are $\Sigma_{cg}$ rather than $\Lambda_{cg}$. Indeed, one can observe in Table~\ref{tab:results} that only one $\Sigma_{cg}$ state is lower than the $\Lambda_{cg}(1/2^+)$ state belonging to the $[\bar{3}_f,\bar{3}_c,0,1,G]$ doublet, while all the others are higher than this $\Lambda_{cg}(1/2^+)$ state.

Additionally, we have examined the charmed hybrid baryons belonging to the other four multiplets---$[6_f,6_c,0,1,G]$, $[\bar{3}_f,6_c,1,0,G]$, $[\bar{3}_f,6_c,1,1,G]$, and $[\bar{3}_f,6_c,1,2,G]$---as well as those with the negative parity. We find that they have significantly larger masses, primarily  because the relevant two-point correlation functions are negative and thus non-physical when $\omega$ is not sufficiently large. 
These results will be discussed in detail in our future studies. 

Before concluding this section, we would like to note that in this calculation, we have focused on the ${\mathcal O}(1/m_Q)$ corrections, particularly those related to the propagators in the QCD sum rules, as described above. However, the ${\mathcal O}(1/m_Q)$ corrections also affect the interpolating currents, known as the flow corrections, whose effects are not considered in the present study. Although the flow corrections are of the same order, their effects are generally smaller and mainly affect the fine details of the coupling constants and Dirac structures. Therefore, they are expected to have a relatively smaller effect compared to the propagator corrections~\cite{Neubert:1993mb,Grozin:1997ih}. For simplicity, we have neglected these corrections in the current analysis, but they will be systematically explored in future studies.

\begin{table*}[hbtp]
\begin{center}
\renewcommand{\arraystretch}{1.6}
\caption{Parameters of the charmed hybrid baryons belonging to the multiplets $[\bar{3}_f,\bar{3}_c,0,1,G]$, $[6_f,\bar{3}_c,1,0,G]$, $[6_f,\bar{3}_c,1,1,G]$, and $[6_f,\bar{3}_c,1,2,G]$, calculated using the QCD sum rule method within the framework of heavy quark effective theory.}
\begin{tabular}{ c | c | c | c | c | c | c | c}
\hline\hline
\multirow{2}{*}{~~~Multiplets~~~} & ~~~~$\omega_c$~~~~ & ~~~~Working region~~~~ & ~~~~~~~~$\overline{\Lambda}$~~~~~~~~ & ~~~~~~\multirow{2}{*}{State}~~~~~~ & ~~~~Mass~~~~ & ~~~~Difference~~~~ & ~~~Decay constant~~~
\\                                               & (GeV) & (GeV)      & (GeV)                &                                    & (GeV)      & (GeV)        & (GeV$^{4}$)
\\ \hline\hline
\multirow{4}{*}{$[\bar{3}_f,\bar{3}_c,0,1,G]$} & \multirow{2}{*}{3.20} & \multirow{2}{*}{$0.400\le T \le 0.410$} & \multirow{2}{*}{$2.72^{+0.17}_{-0.15}$} & $\Lambda_{cg}(1/2^+)$ & $4.74^{+0.23}_{-0.20}$ & \multirow{2}{*}{$0.62^{+0.12}_{-0.10}$} & $0.155^{+0.031}_{-0.024}$
\\ \cline{5-6}\cline{8-8}
  & & & & $\Lambda_{cg}(3/2^+)$ & $
5.36^{+0.30}_{-0.26}$ & &$0.045 ^{+0.009}_{-0.007}$
\\ \cline{2-8}
 & \multirow{2}{*}{3.45} & \multirow{2}{*}{$0.409\le T \le 0.437$} & \multirow{2}{*}{$2.76^{+0.11}_{-0.09}$} & $\Xi_{cg}(1/2^+)$ & $4.91^{+0.15}_{-0.14}$ & \multirow{2}{*}{$0.54^{+0.10}_{-0.08}$} & $0.176^{+0.023}_{-0.019}$
\\ \cline{5-6}\cline{8-8}
  & & & & $\Xi_{cg}(3/2^+)$ & $5.45^{+0.20}_{-0.17}$ & &$0.051^{+0.007}_{-0.005}$
\\ \cline{1-8}
\multirow{3}{*}{$[6_f,\bar{3}_c,1,0,G]$} &$2.75$ & $0.394\leq T \leq 0.396$ & $2.25\pm 0.12$ & $\Sigma_{cg}(1/2^+)$ & $4.75^{+0.21}_{-0.20}$ &--& $0.091^{+0.015}_{-0.013}$
\\ \cline{2-8}
& 3.00 & $0.373\le T \le 0.420$ & $2.30^{+0.14}_{-0.07} $& $\Xi^\prime_{cg}(1/2^+)$ & $4.88^{+0.20}_{-0.13}$ & --& $0.104^{+0.021}_{-0.011}$
\\ \cline{2-8}
 & 3.25 & $0.357\le T \le 0.442$ & $2.41^{+0.15}_{-0.08}$& $\Omega_{cg}(1/2^+)$ & $5.09^{+0.21}_{-0.14}$ & --& $0.183^{+0.041}_{-0.019}$
\\ \hline
\multirow{6}{*}{$[6_f,\bar{3}_c,1,1,G]$} & \multirow{2}{*}{2.95} & \multirow{2}{*}{$0.368\le T \le 0.398$} & \multirow{2}{*}{$2.46^{+0.15}_{-0.14}$} & $\Sigma_{cg}(1/2^+)$ & $3.36^{+0.27}_{-0.26}$ & \multirow{2}{*}{$2.16^{+0.32}_{-0.29}$} & $0.079^{+0.015}_{-0.013}$
\\ \cline{5-6}\cline{8-8}
  & & & & $\Sigma_{cg}(3/2^+)$ & $
5.51^{+0.29}_{-0.26}$ & &$0.046 ^{+0.009}_{-0.007}$
\\ \cline{2-8}
 & \multirow{2}{*}{3.20} & \multirow{2}{*}{$0.351\le T \le 0.423$} & \multirow{2}{*}{$2.50\pm 0.08 $} & $\Xi^\prime_{cg}(1/2^+)$ & $3.60\pm 0.20$ & \multirow{2}{*}{$2.15^{+0.30}_{-0.27}$} & $0.090\pm 0.010$
\\ \cline{5-6}\cline{8-8}
  & & & & $\Xi^\prime_{cg}(3/2^+)$ & $
5.74^{+0.21}_{-0.19}$ & &$0.052\pm 0.006$
\\ \cline{2-8}
 & \multirow{2}{*}{3.45} & \multirow{2}{*}{$0.351\le T \le 0.447$} & \multirow{2}{*}{$2.62\pm 0.08$} & $\Omega_{cg}(1/2^+)$ & $3.82\pm 0.21$ & \multirow{2}{*}{$2.14^{+0.30}_{-0.27}$} & $0.158^{+0.018}_{-0.017}$
\\ \cline{5-6}\cline{8-8}
  & & & & $\Omega_{cg}(3/2^+)$ & $
5.96^{+0.22}_{-0.19}$ & &$0.091^{+0.010}_{-0.010}$
\\ \hline
\multirow{6}{*}{$[6_f,\bar{3}_c,1,2,G]$} & \multirow{2}{*}{3.50} & \multirow{2}{*}{$0.418\le T \le 0.429$} & \multirow{2}{*}{$2.98^{+0.16}_{-0.14}$} & $\Sigma_{cg}(3/2^+)$ & $6.63^{+0.36}_{-0.30}$ & \multirow{2}{*}{$-1.39 ^{+0.21}_{-0.23}$} & $0.364^{+0.063}_{-0.051}$
\\ \cline{5-6}\cline{8-8}
  & & & & $\Sigma_{cg}(5/2^+)$ & $
5.24^{+0.25}_{-0.22}$ & &$0.123 ^{+0.021}_{-0.017}$
\\ \cline{2-8}
 & \multirow{2}{*}{3.75} & \multirow{2}{*}{$0.428\le T \le 0.457$} & \multirow{2}{*}{$3.04^{+0.10}_{-0.09}$} & $\Xi^\prime_{cg}(3/2^+)$ & $6.76^{+0.28}_{-0.24}$ & \multirow{2}{*}{$-1.35^{+0.18}_{-0.20}$} & $0.422^{+0.052}_{-0.043}$
\\ \cline{5-6}\cline{8-8}
  & & & & $\Xi^\prime_{cg}(5/2^+)$ & $
5.42^{+0.17}_{-0.16}$ & &$0.143^{+0.018}_{-0.015}$
\\ \cline{2-8}
 & \multirow{2}{*}{4.00} & \multirow{2}{*}{$0.434\le T \le 0.484$} & \multirow{2}{*}{$3.16 \pm 0.09$} & $\Omega_{cg}(3/2^+)$ & $6.99^{+0.28}_{-0.24}$ & \multirow{2}{*}{$-1.31^{+0.18}_{-0.20}$} & $0.740^{+0.080}_{-0.073}$
\\ \cline{5-6}\cline{8-8}
  & & & & $\Omega_{cg}(5/2^+)$ & $
5.69^{+0.17}_{-0.16}$ & &$0.250^{+0.027}_{-0.025}$
\\ \hline\hline
\end{tabular}
\label{tab:results}
\end{center}
\end{table*}

\section{Discussions and Conclusion}
\label{sec:summary} 

In recent years substantial theoretical and experimental efforts have been devoted to the study of glueballs and hybrid mesons. However, their internal structures remain elusive, underscoring the necessity for continued investigations---particularly into the comparatively less-explored hybrid baryons. In this paper we employ the QCD sum rule method within the framework of heavy quark effective theory to investigate charmed hybrid baryons with various quantum numbers. We systematically construct twenty-eight interpolating currents for these baryons and utilize seven of them to perform numerical analyses. These seven currents form the four baryon multiplets $[\bar{3}_f,\bar{3}_c,0,1,G]$, $[6_f,\bar{3}_c,1,0,G]$, $[6_f,\bar{3}_c,1,1,G]$, and $[6_f,\bar{3}_c,1,2,G]$. Altogether, we compute the masses of nineteen states with quark-gluon configurations $qqcg$, $qscg$, and $sscg$ ($q = u/d$). The resulting mass predictions are summarized in Table~\ref{tab:results}. 

In particular, the charmed hybrid baryons belonging to the doublet $[6_f, \bar{3}c, 1, 1, G]$ comprise six states: $\Sigma_{cg}(1/2^+)$, $\Sigma_{cg}(3/2^+)$, $\Xi^\prime_{cg}(1/2^+)$, $\Xi^\prime_{cg}(3/2^+)$, $\Omega_{cg}(1/2^+)$, and $\Omega_{cg}(3/2^+)$. Among these, the three spin-$1/2$ states are predicted to be the lowest in mass:
\begin{eqnarray}
\nonumber M_{\Sigma_{cg}(1/2^+)}&=& 3.36^{+0.27}_{-0.26}~\rm{GeV}\, ,
\\ M_{\Xi^\prime_{cg}(1/2^+)}&=& 3.59\pm 0.20~\rm{GeV}\, ,
\\ \nonumber M_{\Omega_{cg}(1/2^+)}&=& 3.82\pm 0.21~\rm{GeV}\, .
\end{eqnarray}
It is important to note that all these states have positive parity. Recalling the case of hybrid mesons, the MIT bag model~\cite{Barnes:1982tx,Chanowitz:1982qj} suggested that the lightest supermultiplet states correspond to those with $J^{PC} = 1^{--}$, $0^{-+}$, $1^{-+}$, and $2^{-+}$, all of which have negative parity, suggesting it possible for the lightest hybrid baryons to have positive parity. However, we must acknowledge that the situation for hybrid baryons is quite complex due to the interplay between quark-gluon dynamics and the QCD vacuum structure, so the presence of gluonic excitations in hybrid baryons has the potential to lead to a different parity pattern compared to the naive constituent parton model, in which gluonic degrees of freedom are not explicitly considered. These hybrid states still involve many uncertainties, and thus, extensive further research is needed to fully understand their properties.

Charmed hybrid baryons can decay via the excitation of a color-octet $\bar{q}q$ or $\bar{s}s$ pair from the valence gluon, followed by the recombination of the remaining color-octet clusters---$qqc$, $qsc$, or $ssc$---with the excited $\bar{q}q$ or $\bar{s}s$ pair into color-singlet baryons and mesons. Representative decay channels for the charmed hybrid baryons belonging to the doublet $[6_f,\bar{3}_c,1,1,G]$ are listed in Table~\ref{tab:decay}, providing viable avenues for their experimental identification in future studies.

\begin{table}[H]
\begin{center}
\renewcommand{\arraystretch}{1.5}
\caption{Possible decay channels of the charmed hybrid baryons belonging to the doublet $[6_f,\bar{3}_c,1,1,G]$.}
\begin{tabular}{ c | c  }
\hline\hline
States & Decay patterns 
\\ \hline\hline 
\multirow{1}{*}{$\Sigma_{cg}(1/2^+)$} &
$ND$, $\Sigma D_s$, $N D^*$, $\Sigma D_s^*$, $\Delta D$, $\Sigma^* D_s$
\\ \hline
\multirow{1}{*}{$\Xi^\prime_{cg}(1/2^+)$} & 
$\Lambda D$, $\Xi D_s$, $\Lambda D^*$, $\Xi D_s^*$, $\Xi^* D_s$
\\ \hline 
\multirow{1}{*}{$\Omega_{cg}(1/2^+)$} & 
$\Xi D$, $\Xi D^*$, $\Xi^* D$, $\Omega D_s$, $\Xi^* D^*$ 
\\ \hline\hline
\end{tabular}
\label{tab:decay}
\end{center}
\end{table}

\section*{ACKNOWLEDGMENTS}

We are grateful to Shi-Lin Zhu for the helpful discussions. This project is supported by the China Postdoctoral Science Foundation under Grants No.~2024M750049, the National Natural Science Foundation of China under Grant No.~12075019, and the Fundamental Research Funds for the Central Universities.

\end{document}